\documentclass[aps,amsmath,amssymb,reprint,prd,superscriptaddress,longbibliography]{revtex4-1}  
\usepackage{graphicx}  
\usepackage{dcolumn}   
\usepackage{bm}        
\usepackage{amssymb}   
\usepackage{lineno}

\def\sigeff{\sigma_{\mathrm{eff}}}

\def\nuflub8{\phi^\nu_B}
\def\nuflube7{\phi^\nu_{Be}}

\def\kpc{\,\mathrm{kpc}}

\def\GeV{\,\mathrm{GeV}}

\def\GV{\,\mathrm{GV}}
\def\TV{\,\mathrm{TV}}

\def\pbar{\,\bar{\text{p}}}
\def\He{\,\mathrm{He}}
\def\C{\,\mathrm{C}}
\def\N{\,\mathrm{N}}
\def\O{\,\mathrm{O}}
\def\Li{\,\mathrm{Li}}
\def\Be{\,\mathrm{Be}}
\def\B{\,\mathrm{B}}
\def\A{\,\mathrm{A}}

\def\pbar{\,\bar{\text{p}}}

\def\cpbar{c^{\mathrm{sec}}_{\mathrm{\bar{\text{p}}}}}
\def\che{c^{\mathrm{pri}}_{\mathrm{He}}}
\def\cc{c^{\mathrm{pri}}_{\mathrm{C}}}
\def\cnp{c^{\mathrm{pri}}_{\mathrm{N}}}
\def\co{c^{\mathrm{pri}}_{\mathrm{O}}}
\def\cli{c^{\mathrm{sec}}_{\mathrm{Li}}}
\def\cbe{c^{\mathrm{sec}}_{\mathrm{Be}}}
\def\cb{c^{\mathrm{sec}}_{\mathrm{B}}}
\def\cns{c^{\mathrm{sec}}_{\mathrm{N}}}

\def\cpri{c_{i}^{\mathrm{pri}}}
\def\csec{c_{i}^{\mathrm{sec}}}
\def\Rbr{\,R_{\mathrm{br}}}
\def\kpc{\,\mathrm{kpc}}

\def\GeV{\,\mathrm{GeV}}

\def\GV{\,\mathrm{GV}}
\def\TV{\,\mathrm{TV}}

\def\pbar{\,\bar{\text{p}}}
\def\He{\,\mathrm{He}}
\def\C{\,\mathrm{C}}
\def\N{\,\mathrm{N}}
\def\O{\,\mathrm{O}}
\def\Li{\,\mathrm{Li}}
\def\Be{\,\mathrm{Be}}
\def\B{\,\mathrm{B}}
\def\A{\,\mathrm{A}}

\def\pbar{\,\bar{\text{p}}}

\def\cpbar{c^{\mathrm{sec}}_{\mathrm{\bar{\text{p}}}}}
\def\che{c^{\mathrm{pri}}_{\mathrm{He}}}
\def\cc{c^{\mathrm{pri}}_{\mathrm{C}}}
\def\cnp{c^{\mathrm{pri}}_{\mathrm{N}}}
\def\co{c^{\mathrm{pri}}_{\mathrm{O}}}
\def\cli{c^{\mathrm{sec}}_{\mathrm{Li}}}
\def\cbe{c^{\mathrm{sec}}_{\mathrm{Be}}}
\def\cb{c^{\mathrm{sec}}_{\mathrm{B}}}
\def\cns{c^{\mathrm{sec}}_{\mathrm{N}}}

\def\cpri{c_{i}^{\mathrm{pri}}}
\def\csec{c_{i}^{\mathrm{sec}}}

\def\Rbr{\,R_{\mathrm{br}}}

\def\chidof{\,\chi^{2}/\mathrm{d.o.f}}

\def\aap{Astron.\ Astrophys.\ }
\def\apj{Astrophys.\ J.\ }
\def\apjl{Astrophys.\ J.\ Lett.\ }

\def\mnras{Mon.\ Not.\ Roy.\ Astron.\ Soc.\ }

\def\prd{Phys.\ Rev.\ D\ }

\def\jcap{J.\ Cosmol.\ Astropart.\ Phys.\ }
\def\ssr{Space\ Sci.\ Rev.\ }

\begin{document}



\title{Some new hints on cosmic-ray propagation from AMS-02 nuclei spectra}

\author{Jia-Shu Niu}
\email{jsniu@sxu.edu.cn}
\affiliation{Institute of Theoretical Physics, Shanxi University, Taiyuan, 030006, China}

\author{Hui-Fang Xue}%
\affiliation{Department of Physics, Taiyuan Normal University, Taiyuan, 030619, China}



\begin{abstract}
   In this work, we considered 2 schemes (a high-rigidity break in primary source injections and a high-rigidity break in diffusion coefficient) to reproduce the newly released AMS-02 nuclei spectra (He, C, N, O, Li, Be, and B) when the rigidity larger than 50 GV. The fitting results show that current data set favors a high-rigidity break at $\sim 325 \GV$ in diffusion coefficient rather than a break at $\sim 365 \GV$ in primary source injections. Meanwhile, the fitted values of the factors to rescale the cosmic-ray (CR) flux of secondary species/components after propagation show us that the secondary flux are underestimated in current propagation model. It implies that we might locate in a slow diffusion zone, in which the CRs propagate with a small value of diffusion coefficient compared with the averaged value in the galaxy. Another hint from the fitting results show that extra secondary CR nuclei injection may be needed in current data set. All these new hints should be paid more attention in future research.
\end{abstract}

\maketitle


\section{Introduction}

Cosmic-ray (CR) physics has entered a precision-driven era (see, e.g., \citet{Niu2018}). More and more fine structures have been revealed by a new generation of space-borne and ground-based experiments in operation. For nuclei spectra, the most obvious fine structure is the spectral hardening at $\sim$ 300 GV, which was observed by ATIC-2 \citep{ATIC2006}, CREAM \citep{CREAM2010}, PAMELA \citep{PAMELA2011}, and AMS-02 \citep{AMS02_proton,AMS02_helium}.

Recently released AMS-02 nuclei spectra have confirmed that the spectral hardening exists not only in the primary CR nuclei species (He, C, and O \citep{AMS02_He_C_O}), but also in the secondary CR nuclei species (Li, Be, and B \citep{AMS02_Li_Be_B}) and hybrid nuclei species (N\footnote{In CR physics, nitrogen spectrum is thought to contain both primary and secondary components.} \citep{AMS02_N}). This provides us an excellent opportunity to study the spectral hardening and the physics behind it quantitatively.

Some solutions are proposed to explain this observed phenomenon: (i) adding a new break in high-rigidity region ($\sim 300 \GV$) to the primary source injection spectra (see, e.g., \citet{Korsmeier2016,Boschini2017,Niu2018_dampe,Niu2019_dampe,Zhu2018,Niu201810}); (ii) adding a new high-rigidity break in the diffusion coefficient (see, e.g., \citet{Genolini2017,Niu201810}); (iii) inhomogeneous diffusion (see, e.g., \citet{Blasi2012,Tomassetti2012,Tomassetti2015apjl01,Tomassetti2015prd,Feng2016,Guo2018}); (iv) the superposition of local and distant sources (see, e.g., \citet{Vladimirov2012,Bernard2013,Thoudam2013,Tomassetti2015apjl02,Kachelriess2015,Kawanaka2018}).

The above explanations could be divided into two classes at first step: (i), (ii), and (iii), which ascribe the spectral hardening to non-local source effects; (iv), which ascribes it to the contribution of local sources. At second step, the first class of the first step could also be divided into two sub-classes: (i), which ascribes the spectral hardening to the primary source injections; (ii) and (iii), which ascribe it to the propagation processes.  If the spectral hardening comes from the CR sources, the ratio between secondary and primary species' spectra should appear featureless (or the primary and secondary spectra are equally hardened), since the secondary CR spectra inherit the features from the primary CR spectra. On the other hand, if the spectral hardening is due to the propagation processes, the ratio between secondary and primary species' spectra should be featured because the secondary species spectra not only inherit the hardening from the primary species (which is caused by the propagation of primary species), but are also hardened by their own propagation processes. This lead to a harder ''tail'' in CR secondary nuclei spectra than previous case (see, e.g., Fig. 2 in \citet{Niu201810}). In this sense, (ii) and (iii) should have a similar prediction on the spectral hardening in primary and secondary nuclei spectra (see, e.g., \citet{Feng2016,Niu201810}).

 As a result, in this work, we design 2 schemes to test the origin of the spectra hardening in CR nuclei spectra: (a) the spectral hardening comes from the sources, which can be described by a high-rigidity break in the primary source injections (Scheme I); (b) the spectral hardening comes from the propagation processes, which can be described by a high-rigidity break in the diffusion coefficient (Scheme II).  Both of the schemes are implemented by the public code {\sc galprop} v56\footnote{http://galprop.stanford.edu} to reproduce the AMS-02 nuclei spectra in the global fitting. We hope that the AMS-02 data could give us a clear quantitative evidence to the origin of the spectral hardening in CR nuclei species.

\section{Setups}

 As the framework has been established in our previous works \citep{Niu2018,Niu2018_dampe,Niu2019_dampe,Niu201810}, we employ a Markov Chain Monte Carlo (MCMC) algorithm \footnote{based on the {\sc python} module {\tt emcee} (http://dan.iel.fm/emcee/)} which is embedded by {\sc galprop} to do global fitting. The diffusion-reacceleration model is used as the unique propagation model in this work. A uniform diffusion coefficient which depends on CR particles' rigidity is used in the whole propagation region. The propagation region is assumed to have a cylindrical symmetry and a free escape boundary condition. The radial ($r$) and vertical ($z$) grid steps are chosen as $\Delta r = 1 \kpc$, and $\Delta z = 0.2 \kpc$. The grid in kinetic energy per nucleon is logarithmic between $1 \GeV$ and $10^{4} \GeV$ with a step factor of 1.2.  The nuclear network used in our calculations is extended to silicon-28.

Some of the most important setups which are different from our previous work \citep{Niu201810} are listed as follows (more detailed similar configurations could be found in \citet{Niu2018,Niu201810}):
\begin{itemize}
\item[(1)] In this work, we do not use the proton and antiproton spectra in our
  global fitting. On the one hand, there is an obvious difference observed in the slopes of proton and other nuclei species when $Z > 1$ \citep{AMS02_proton,AMS02_helium,AMS02_He_C_O}, which might indicate a different origin of the spectral hardening between proton and other primary CR species and needs to study independently. On the other hand, the spectrum of antiproton is dominately determined by proton spectrum, and might include some extra sources (like dark matter, see, e.g., \citet{Cui2017,Cuoco2017}). Excluding these data would help us to focus on the main aims of this work and avoid some unknown bias in the global fitting.
\item[(2)] The CR nuclei spectra are seriously influenced by solar modulation when the rigidity below 30 - 40 GV. Moreover, \citet{AMS02_time_proton_he} has proved that the CR spectra of proton and helium are varying in solar cycle 24 when $R \lesssim 40 \GV$. At the same time, in AMS-02 nuclei spectra, data points from 1 GV to 30 GV always have small uncertainties, which seriously influence the global fitting results. Consequently, we use the data points above 50 GV to do the global fitting, which could avoid the influences from low-rigidity data points and solar modulation model, and concentrate on the spectral hardening in high-rigidity region.
\item[(3)] Some of the free parameters which are not directly related to the high-rigidity spectra are removed or fixed as the best-fit values in our previous work \citep{Niu201810}. In detail, the low-rigidity slopes and breaks in primary source injections,  all the solar modulation potentials ($\phi_{i}$s), and all the parameters directly related to proton and antiproton spectra are removed. $D_{0}$ (the normalization of the diffusion coefficient), $z_{h}$ (the half-height of the propagation region), and  $v_{A}$ (the Alfven velocity) are fixed.
\item[(4)] In this work, all the nuclei spectra data in the global fitting comes from AMS-02, which could avoid the complicities to combine the systematics from different experiments.
\item[(5)] In this work, the nitrogen spectrum (which is thought to be contributed both by primary and secondary components) is employed in the global fitting.
\end{itemize}

Altogether, the data set in our global fitting is
\begin{align*}
   \boldsymbol{D} = &\{D^{\text{AMS-02}}_{\He},  D^{\text{AMS-02}}_{\C}, D^{\text{AMS-02}}_{\N}, D^{\text{AMS-02}}_{\O}, \\
   &D^{\text{AMS-02}}_{\Li}, D^{\text{AMS-02}}_{\Be},  D^{\text{AMS-02}}_{\B}  \}~.
 \end{align*}

 In Scheme I, the diffusion coefficient is parametrized as
\begin{equation}
\label{eq:diffusion_coefficient_1}
D_{xx}(R) = D_0\beta \left( \frac{R}{R_{0}} \right)^{\delta}~,~\,
\end{equation}
where $\beta$ is the velocity of the particle in unit of light speed $c$, $R\equiv pc/Ze$ is the rigidity of a particle, and $R_{0}$ is the reference rigidity (4 GV).

For Scheme II, the diffusion coefficient is parametrized as
\begin{equation}
\label{eq:diffusion_coefficient_2}
D_{xx}(R)=  D_{0} \cdot \beta \left(\frac{\Rbr}{R_{0}}\right) \times \left\{
  \begin{array}{ll}
    \left( \dfrac{R}{\Rbr} \right)^{\delta_{1}} & R \le \Rbr\\
    \left( \dfrac{R}{\Rbr} \right)^{\delta_{2}} & R > \Rbr
  \end{array}
  \right.,
\end{equation}
where $\Rbr$ is the high-rigidity break, $\delta_{1}$ and $\delta_{2}$ are the diffusion slopes below and above the break.

The primary source injection spectra of all kinds of nuclei are assumed to be a broken power law form. In Scheme I, it is represented as:
\begin{equation}
  q_{\mathrm{i}} =  N_{\mathrm{i}} \times \left\{ \begin{array}{ll}
\left( \dfrac{R}{R\mathrm{_{A}}} \right)^{-\nu_{\A1}}  & R \leq R_{\A}\\
\left( \dfrac{R}{R\mathrm{_{A}}} \right)^{-\nu_{\A2}} & R > R_{\A}
  \end{array}
  \right.,
  \label{eq:injection_spectra_1}
\end{equation}
where $\mathrm{i}$ denotes the species of nuclei, $N_\mathrm{i}$ is the normalization constant proportional to the relative abundance of the corresponding nuclei, and $\nu_{\A1} / \nu_{\A2}$ for the nucleus rigidity $R$ in the region divided by the break at the high-rigidity $R_{\A}$. In this work, all the nuclei are assumed to have the same value of injection parameters.

For Scheme II, we have
\begin{equation}
  q_{\mathrm{i}} =  N_{\mathrm{i}} \times R^{-\nu_{\A}}
  \label{eq:injection_spectra_2}
\end{equation}
which are described by a power law with an index $\nu_{\A}$.

 In {\sc galprop}, the primary source isotopic abundances are determined by fitting to the data from ACE at about 200 MeV/nucleon, based on a special propagation model \citep{Wiedenbeck2001,Wiedenbeck2008}. But this appears some discrepancies when fit to some new data (see, e.g., \citet{Johannesson2016}), which covers high-energy regions. Consequently, in both of the 2 schemes, $\che$, $\cc$, $\cnp$, and $\co$ are employed to rescale the default abundances of helium-4 ($7.199 \times 10^{4}$), carbon-12 ($2.819 \times 10^{3}$), nitrogen-14 ($1.828 \times 10^{2}$), and oxygen-16 ($3.822 \times 10^{3}$). \footnote{In {\sc galprop}, the abundance of proton is fixted to be a value of $10^{6}$, and all the values in the parenthesis represent the relative abundances to that of proton.}

 At the same time, some works (see, e.g., \citet{Niu2018,Niu201810}) show that if one wants to fit the CR secondary spectra successfully, one should employ factors to rescale the flux of them after propagation. For antiproton, this factor is always in the region 0.8-1.9 \citep{Lin2015,Lin2017,Niu2018}, and always interpreted as the uncertainties from the antiproton production cross section \citep{Tan1983,Duperray2003,Kappl2014,Mauro2014}. In our previous work \citep{Niu201810}, we found that all these factors are systematically larger than 1.0. This confirmed the necessity to employ them in the global fitting and lead us to find the physics behind them. Consequently, in this work, $\cli$, $\cbe$, $\cb$, and $\cns$ are employed to rescale the flux of the secondary CR nuclei species (Li, Be, and B), and the secondary component of N.

In summary, the parameter set for Scheme I is
\begin{align*}
  \boldsymbol{\theta}_{1} =  &\{\delta, R_{\A}, \nu_{\A1}, \nu_{\A2}, | \\
                             & \che, \cc, \cnp, \co, | \\
                             & \cli, \cbe, \cb, \cns \}~,
\end{align*}
for Scheme II is
\begin{align*}
  \boldsymbol{\theta}_{2} =  &\{\Rbr,  \delta_{1}, \delta_{2}, \nu_{\A}, | \\
                             & \che, \cc, \cnp, \co, | \\
                             & \cli, \cbe, \cb, \cns \}~.
\end{align*}

\section{Fitting Results}

\begin{figure*}[!htbp]
  \centering
  \includegraphics[width=0.495\textwidth]{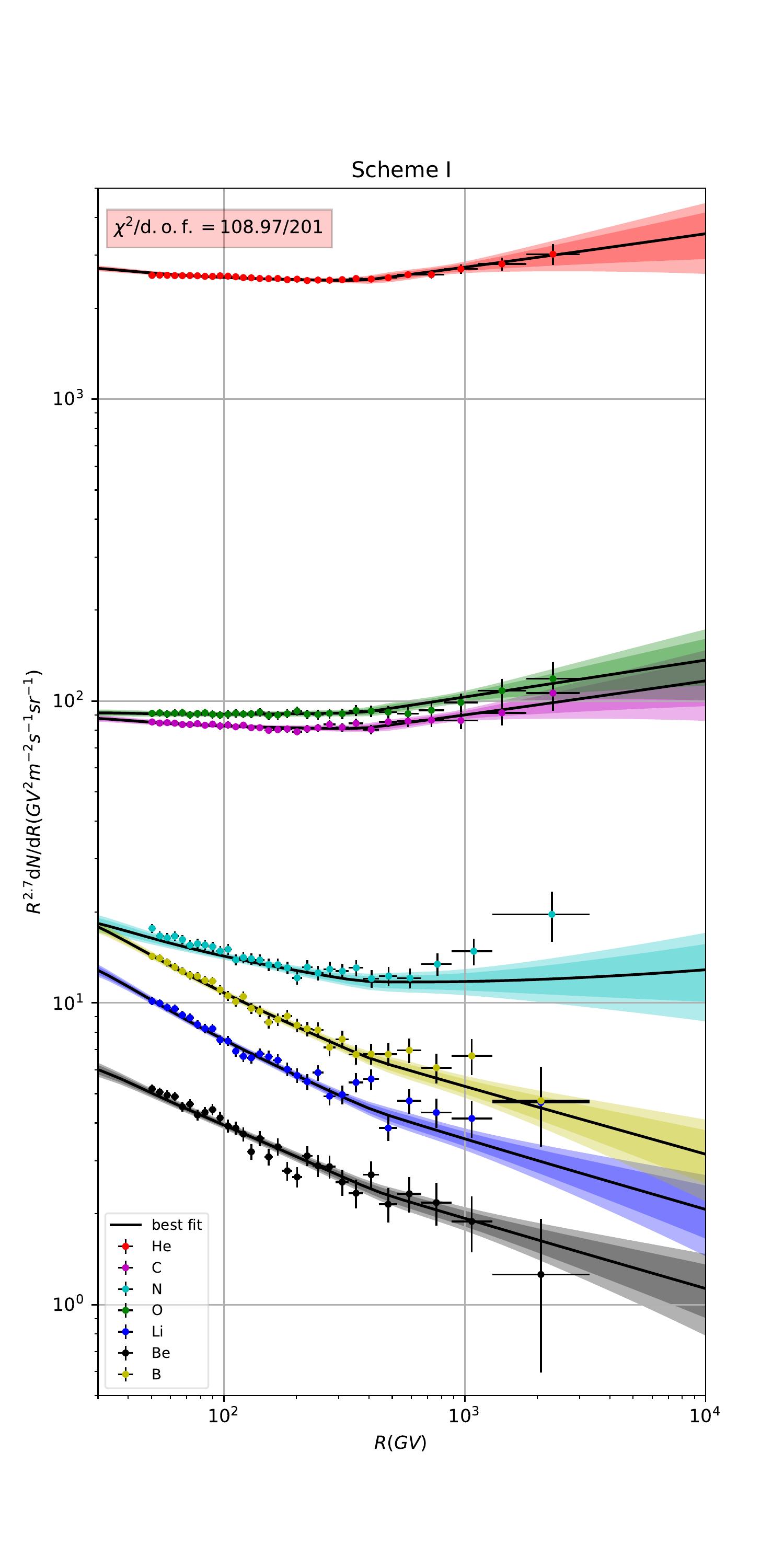}
  \includegraphics[width=0.495\textwidth]{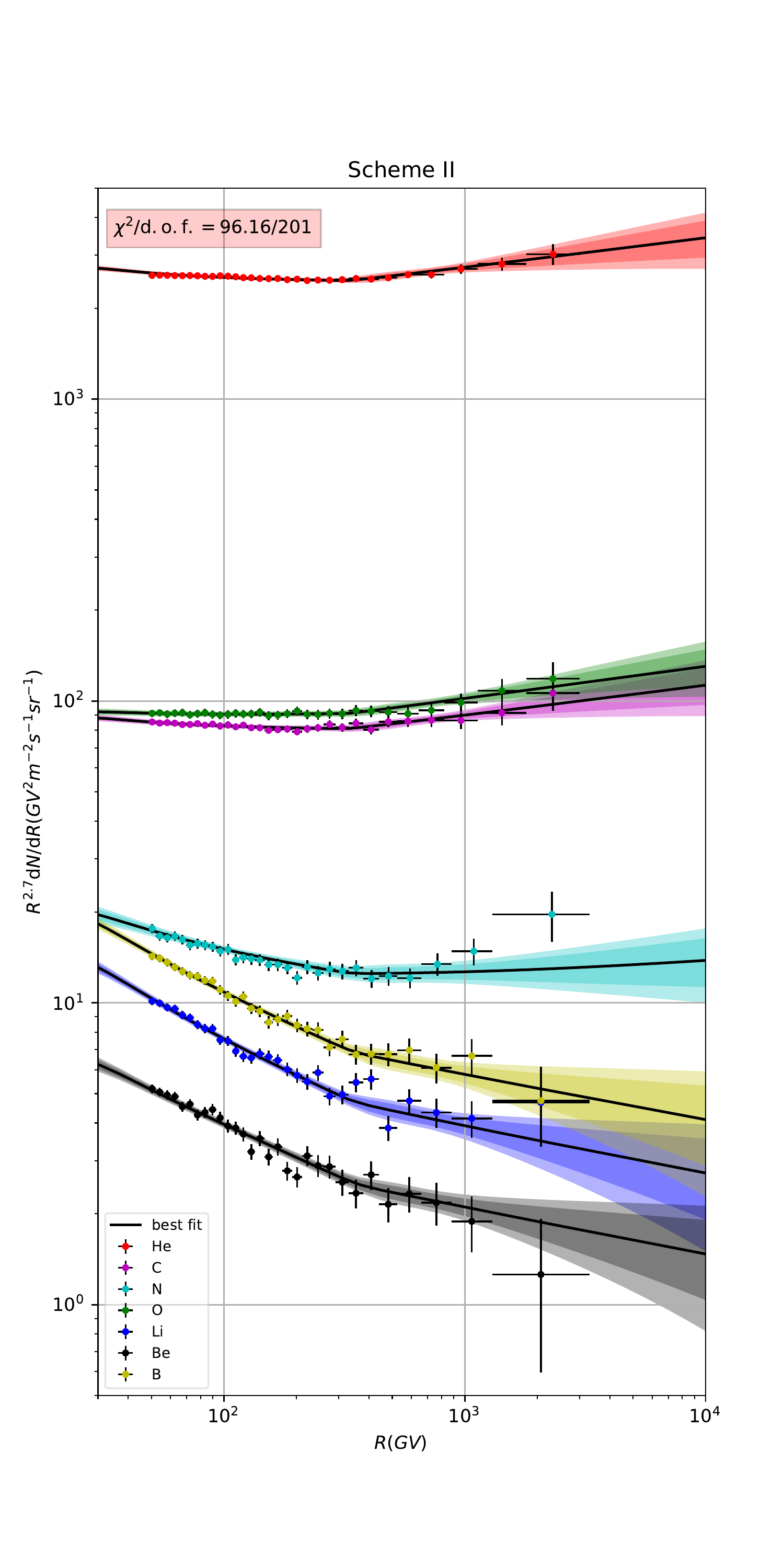}
  \caption{The global fitting results of all the CR nuclei species employed in this work for Scheme I and II. The $2\sigma$ (deeply colored) and $3\sigma$ (lightly colored) bounds are also shown in the sub-figures. The relevant $\chidof$ of the two schemes are also given in the sub-figures.}
\label{fig:all_nuclei_results}
\end{figure*}

\begin{figure*}[!htbp]
  \centering
  \includegraphics[width=0.495\textwidth]{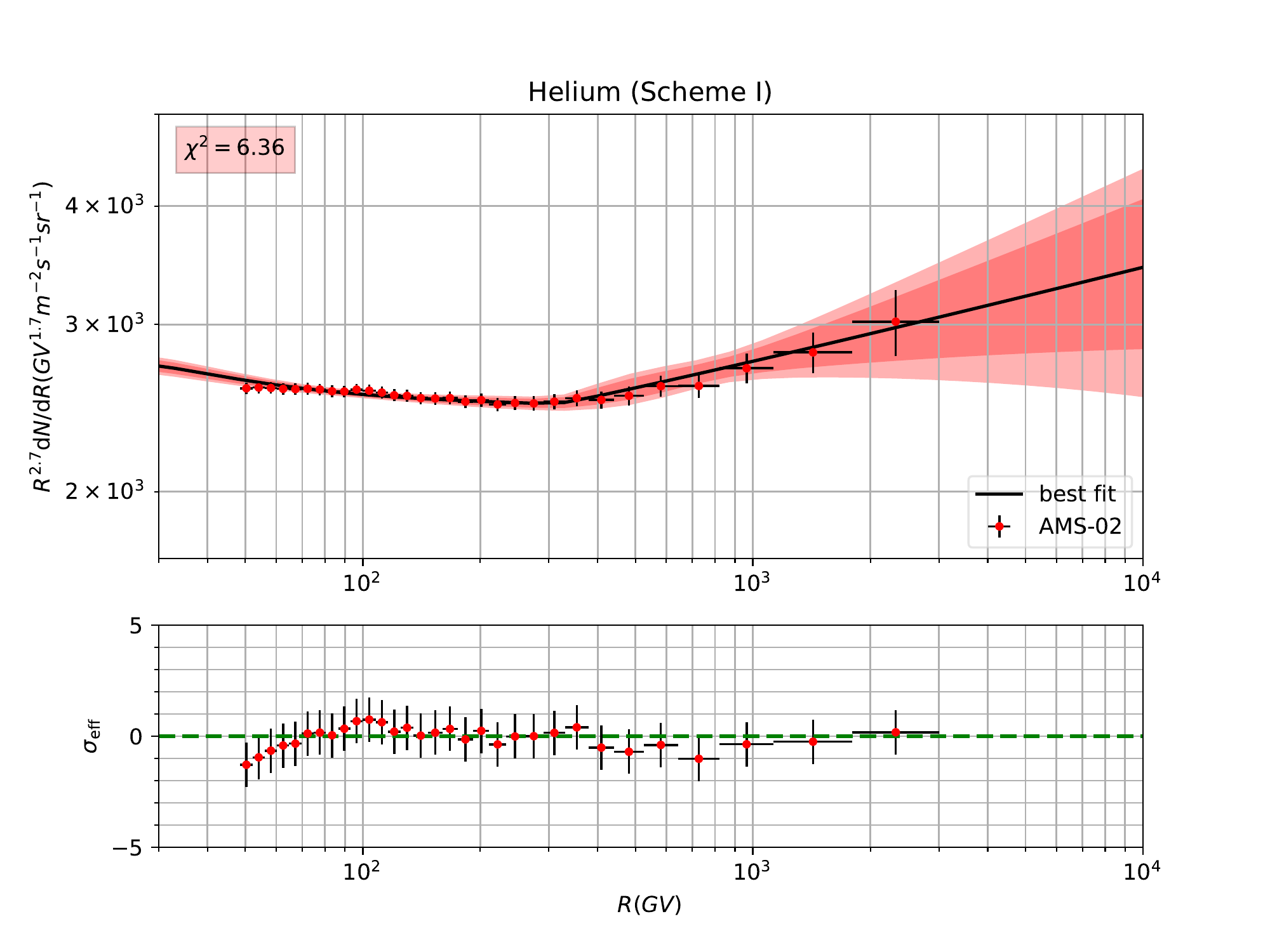}
  \includegraphics[width=0.495\textwidth]{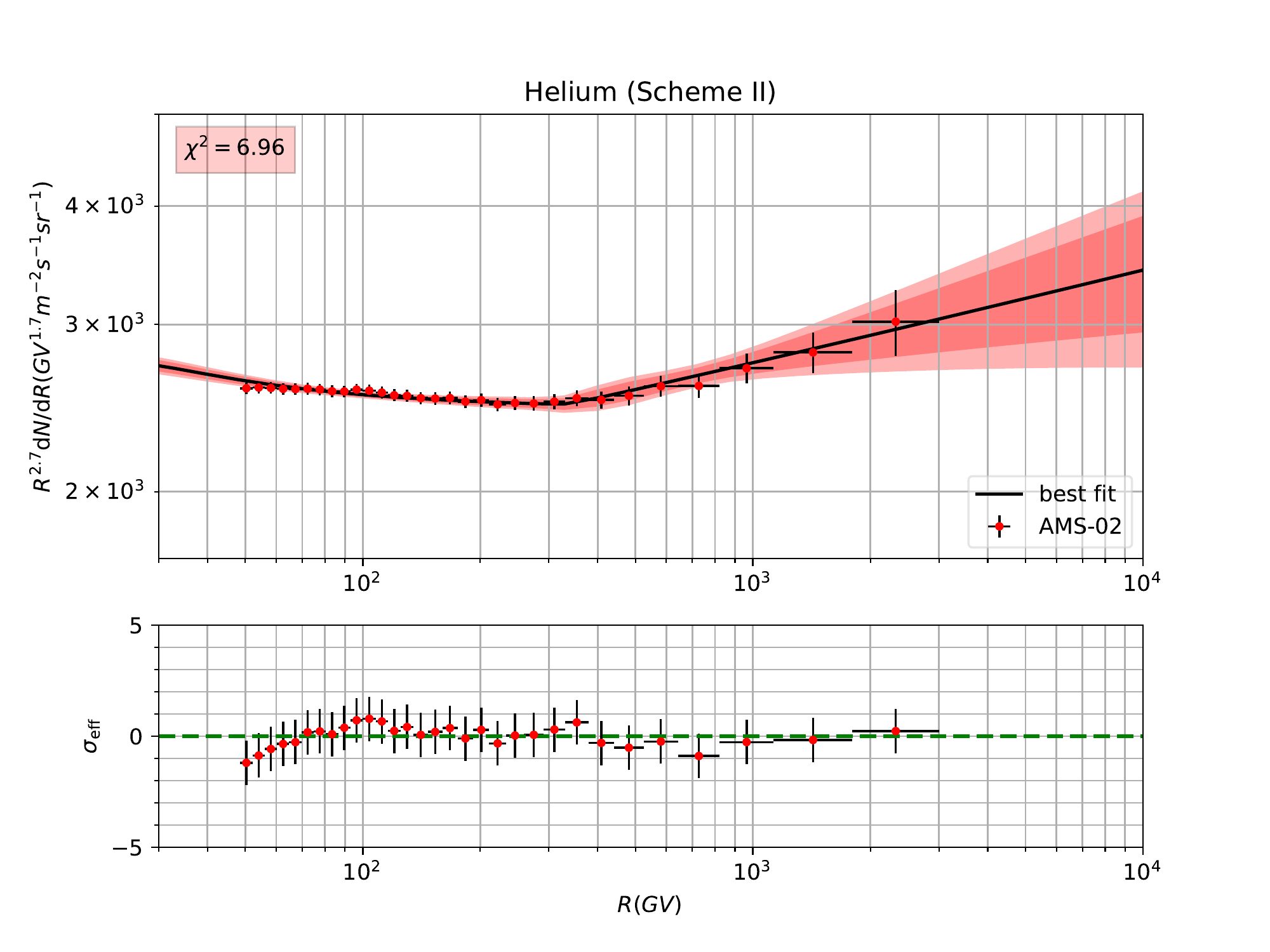}
  \includegraphics[width=0.495\textwidth]{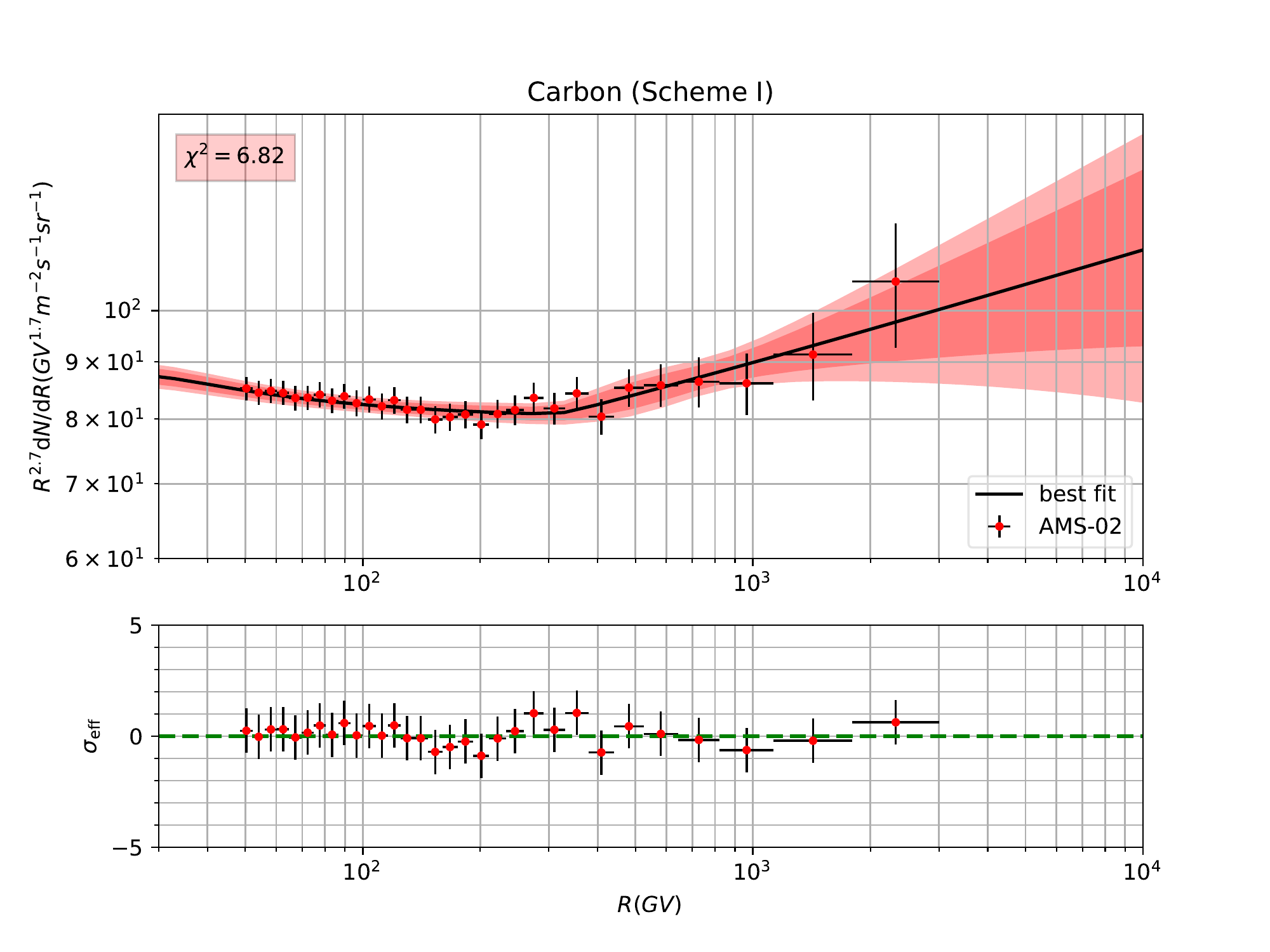}
  \includegraphics[width=0.495\textwidth]{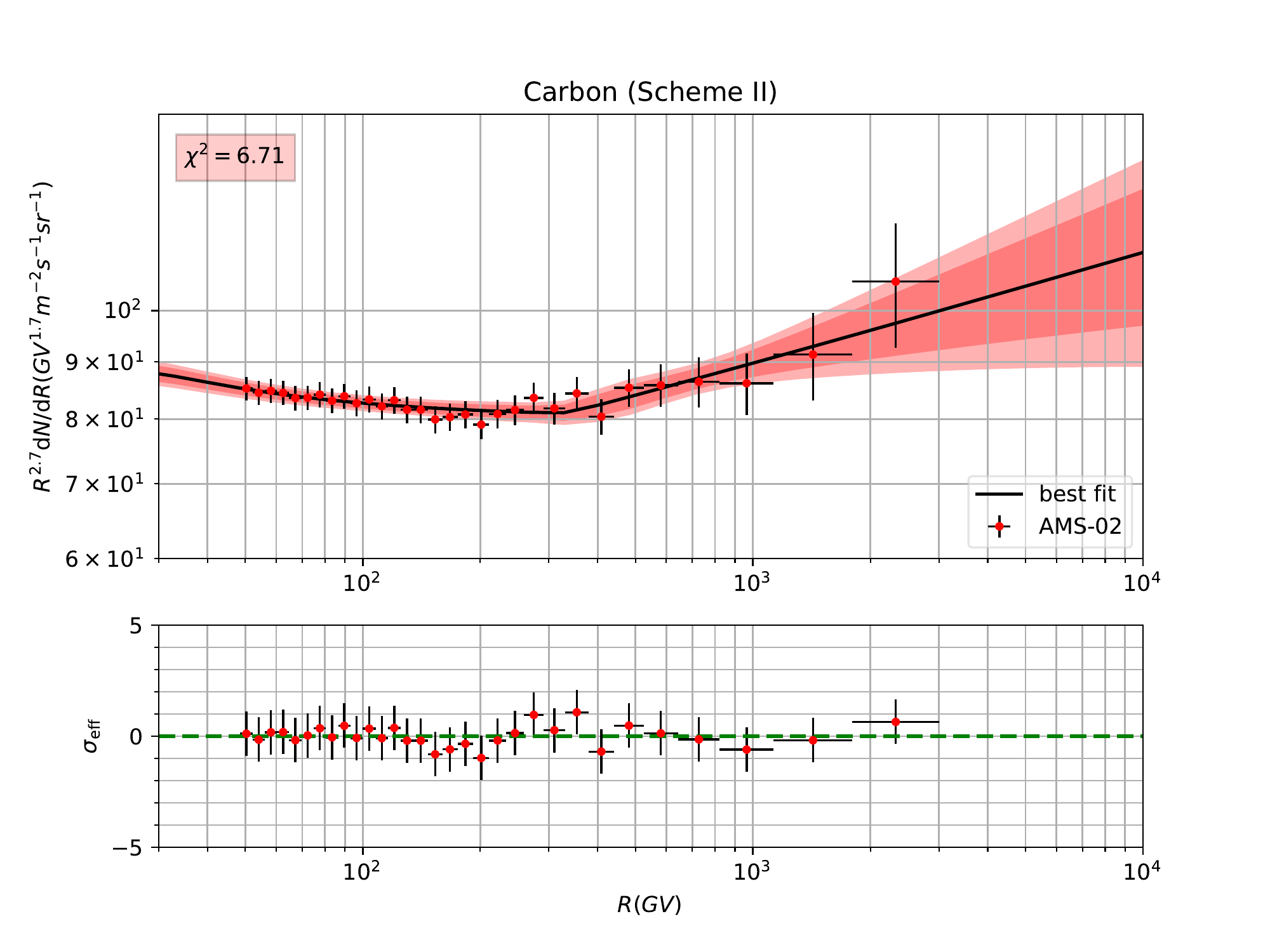}
  \includegraphics[width=0.495\textwidth]{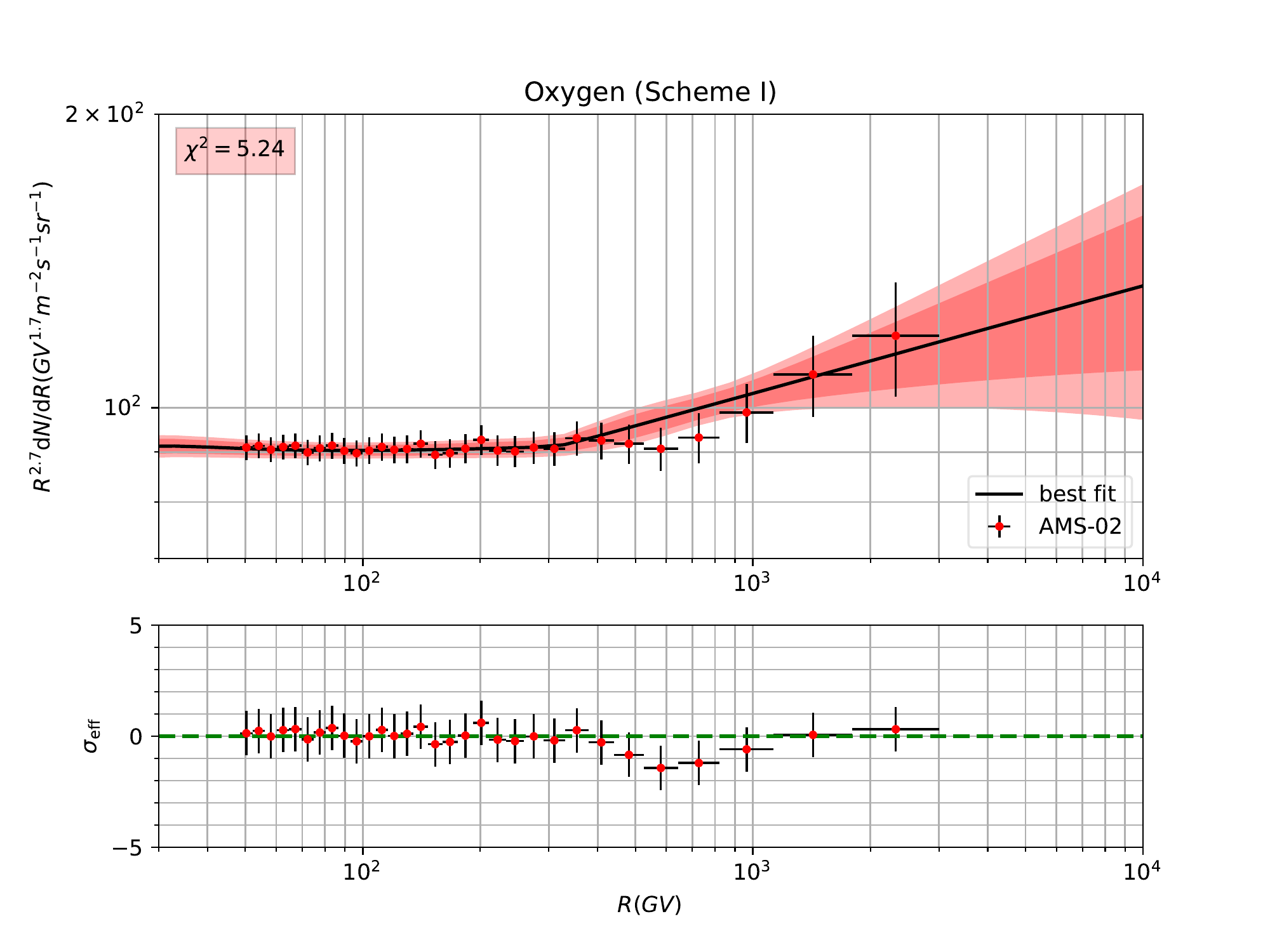}
  \includegraphics[width=0.495\textwidth]{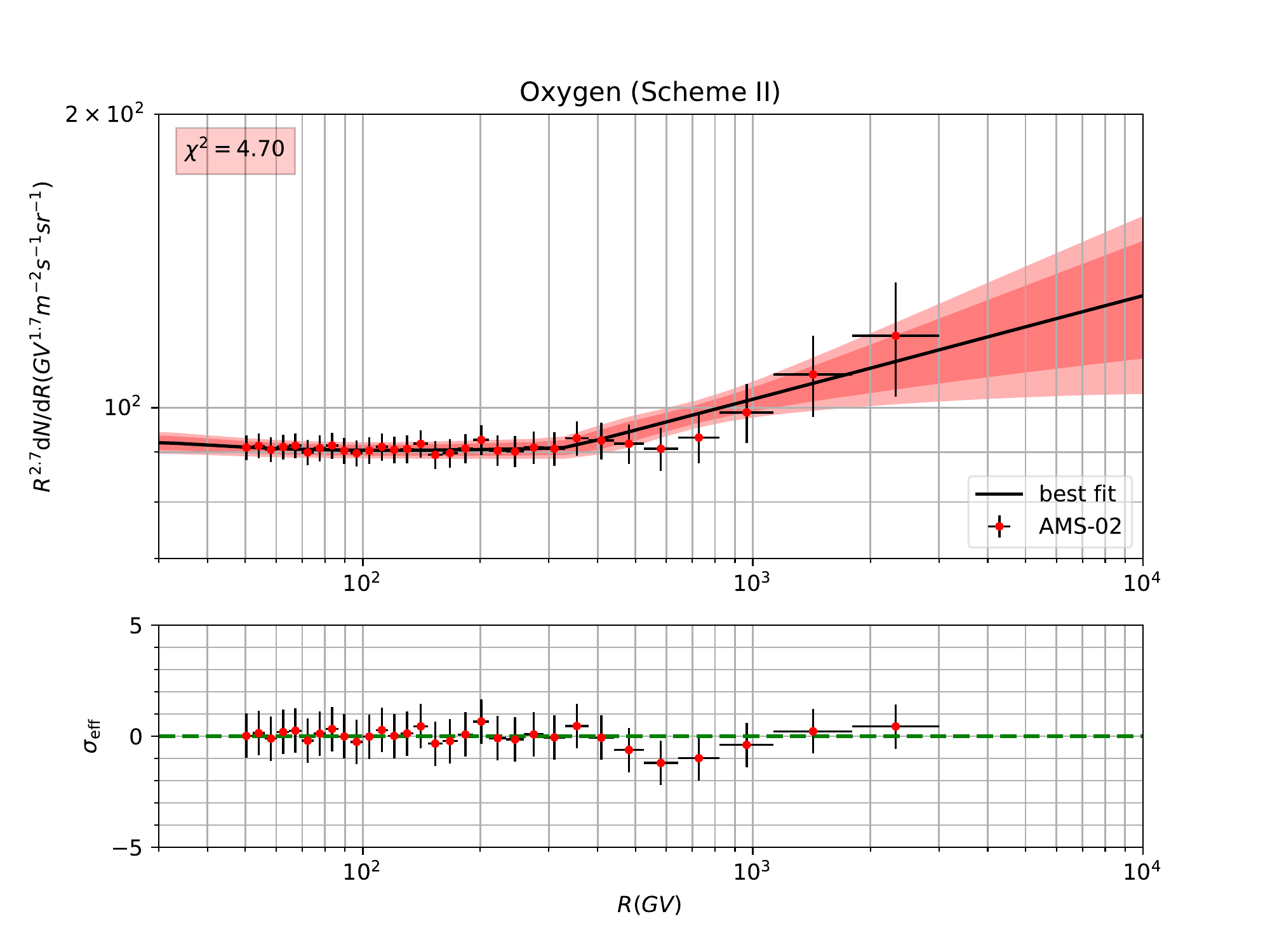}
  \caption{The global fitting results of the primary CR nuclei spectra (He, C,  and O) for two schemes. The $2\sigma$ (deep red) and $3\sigma$ (light red) bounds are also shown in the sub-figures. The relevant $\chi^{2}$ of each nuclei species is given in the sub-figures as well.}
\label{fig:primary_results}
\end{figure*}

\begin{figure*}[!htbp]
  \centering
  \includegraphics[width=0.495\textwidth]{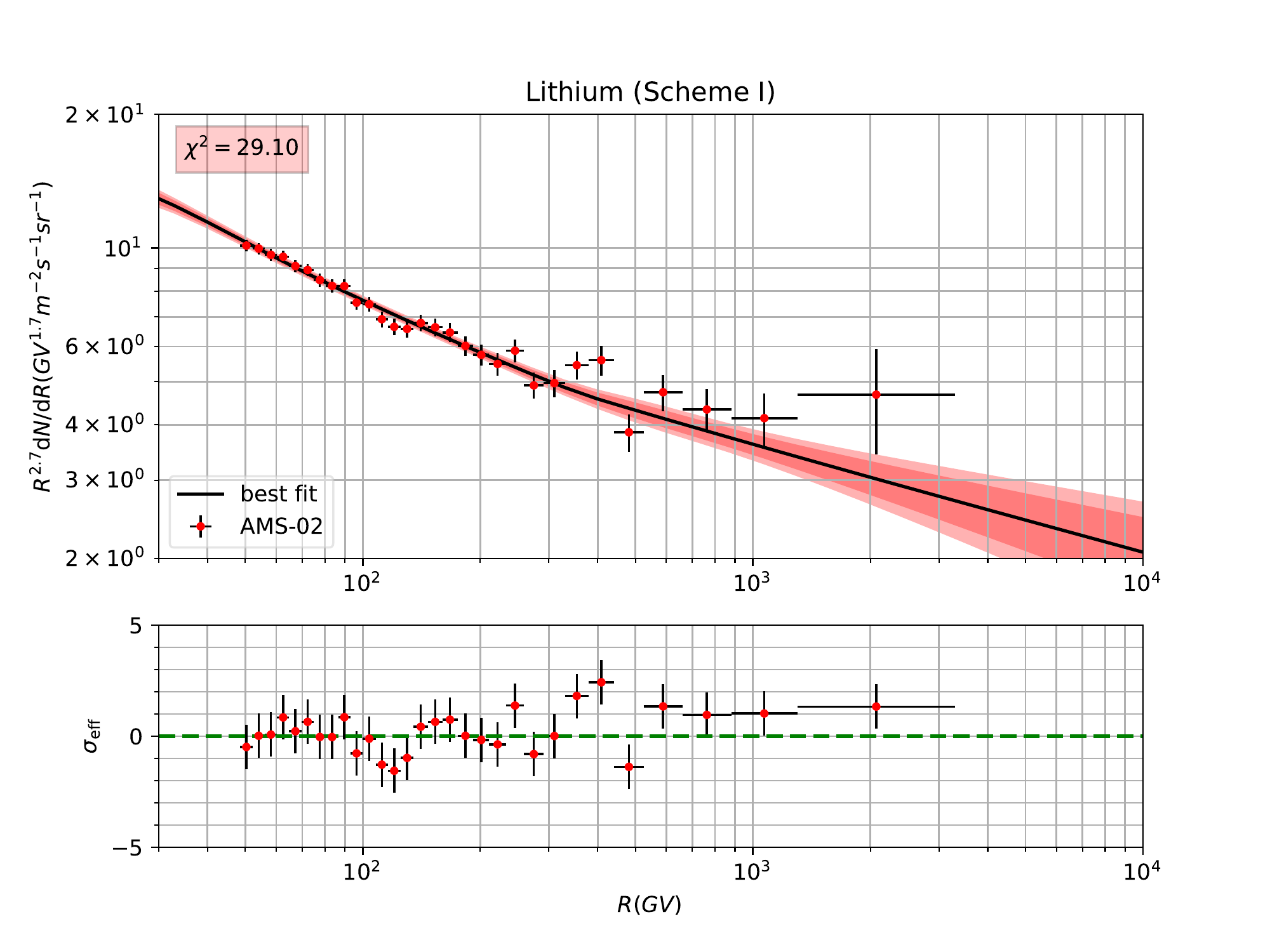}
  \includegraphics[width=0.495\textwidth]{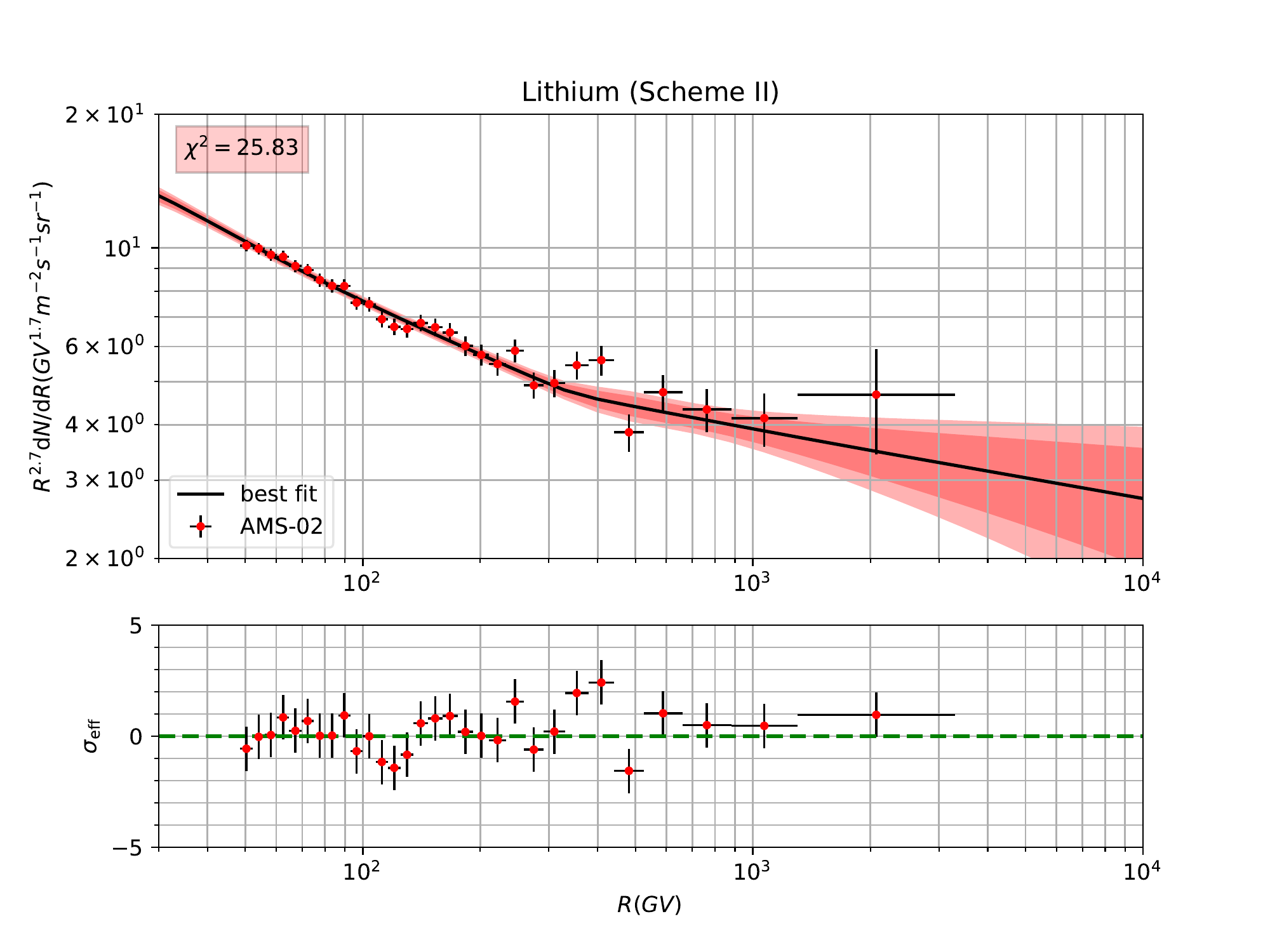}
  \includegraphics[width=0.495\textwidth]{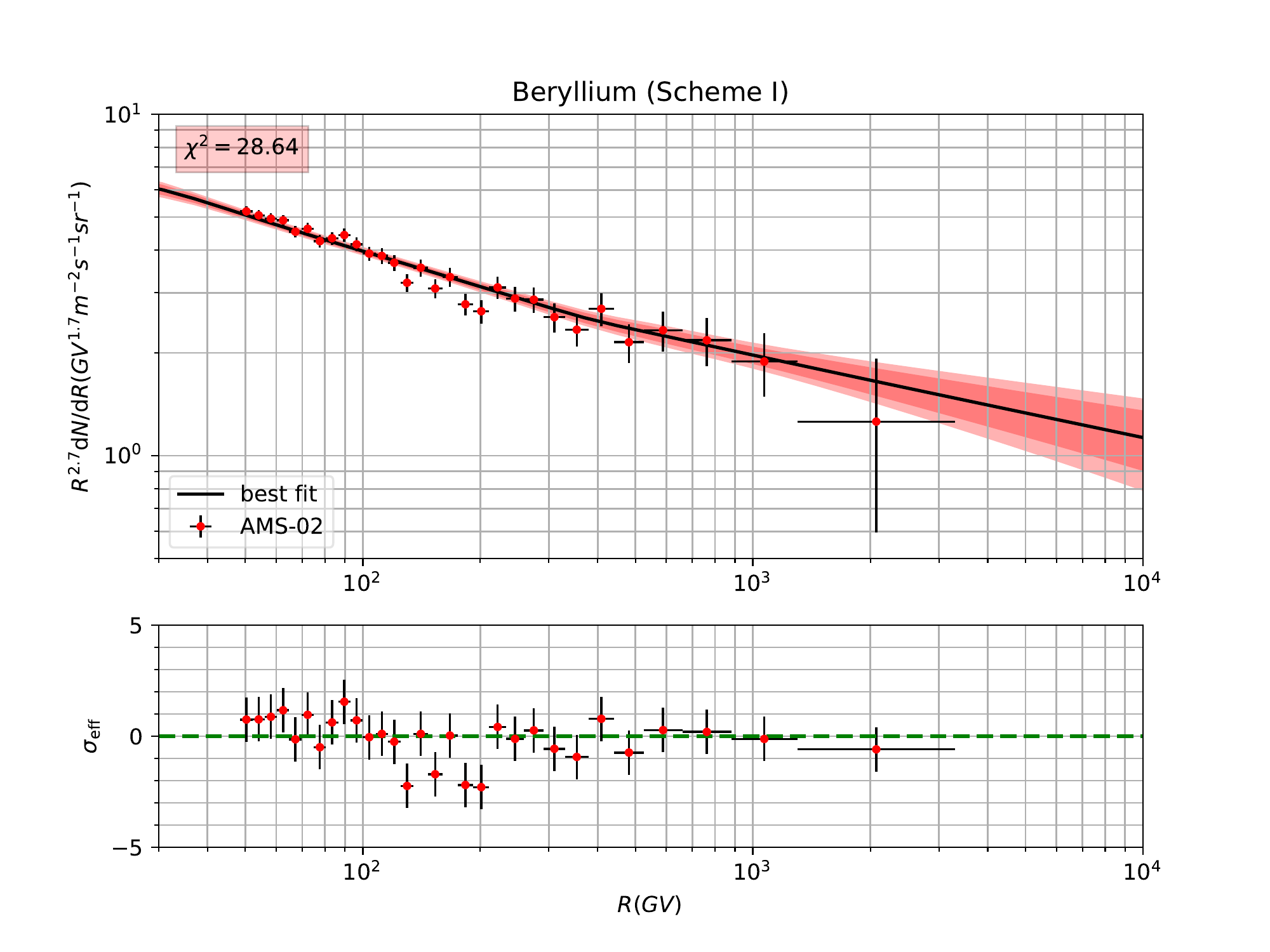}
  \includegraphics[width=0.495\textwidth]{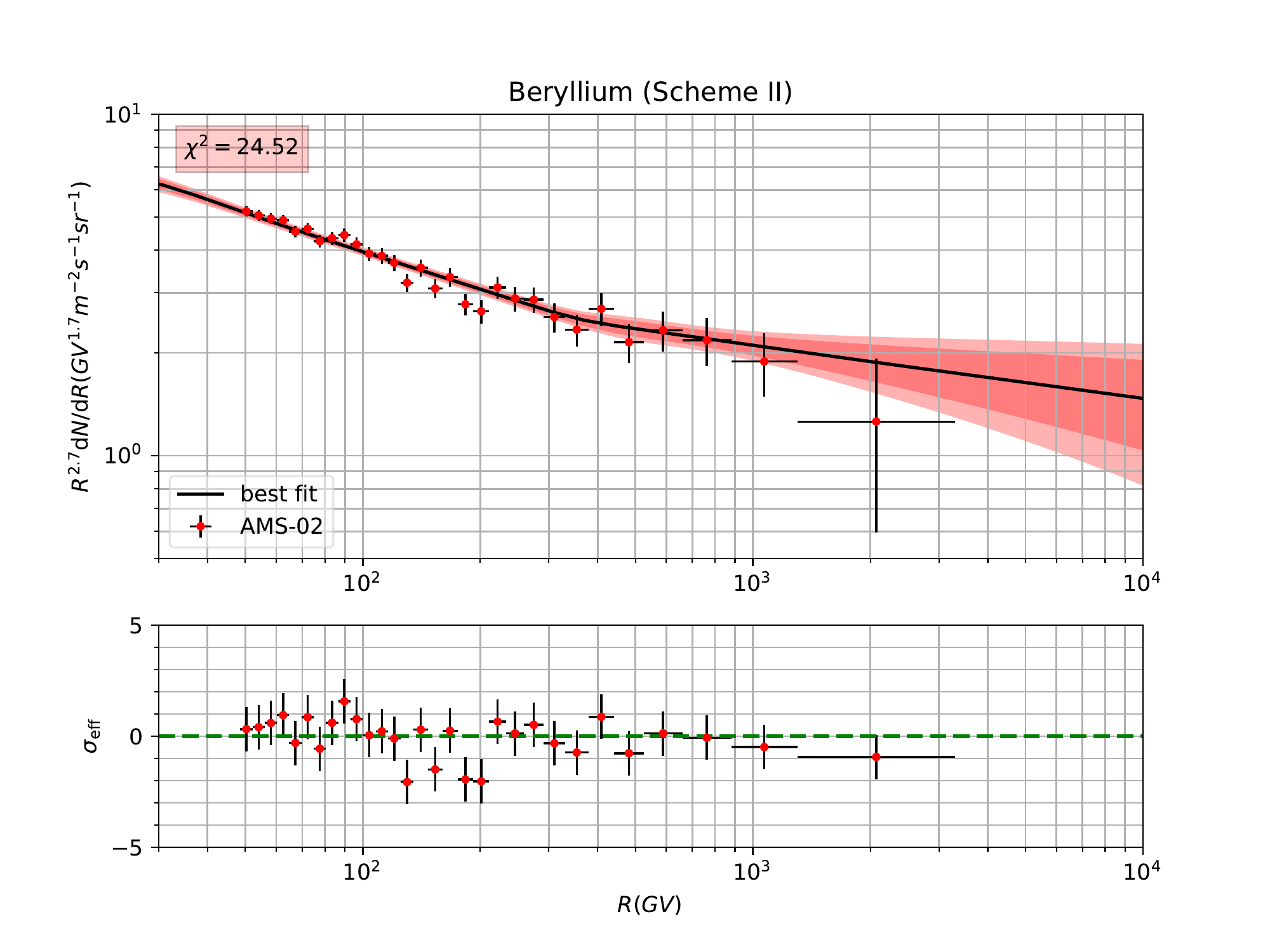}
  \includegraphics[width=0.495\textwidth]{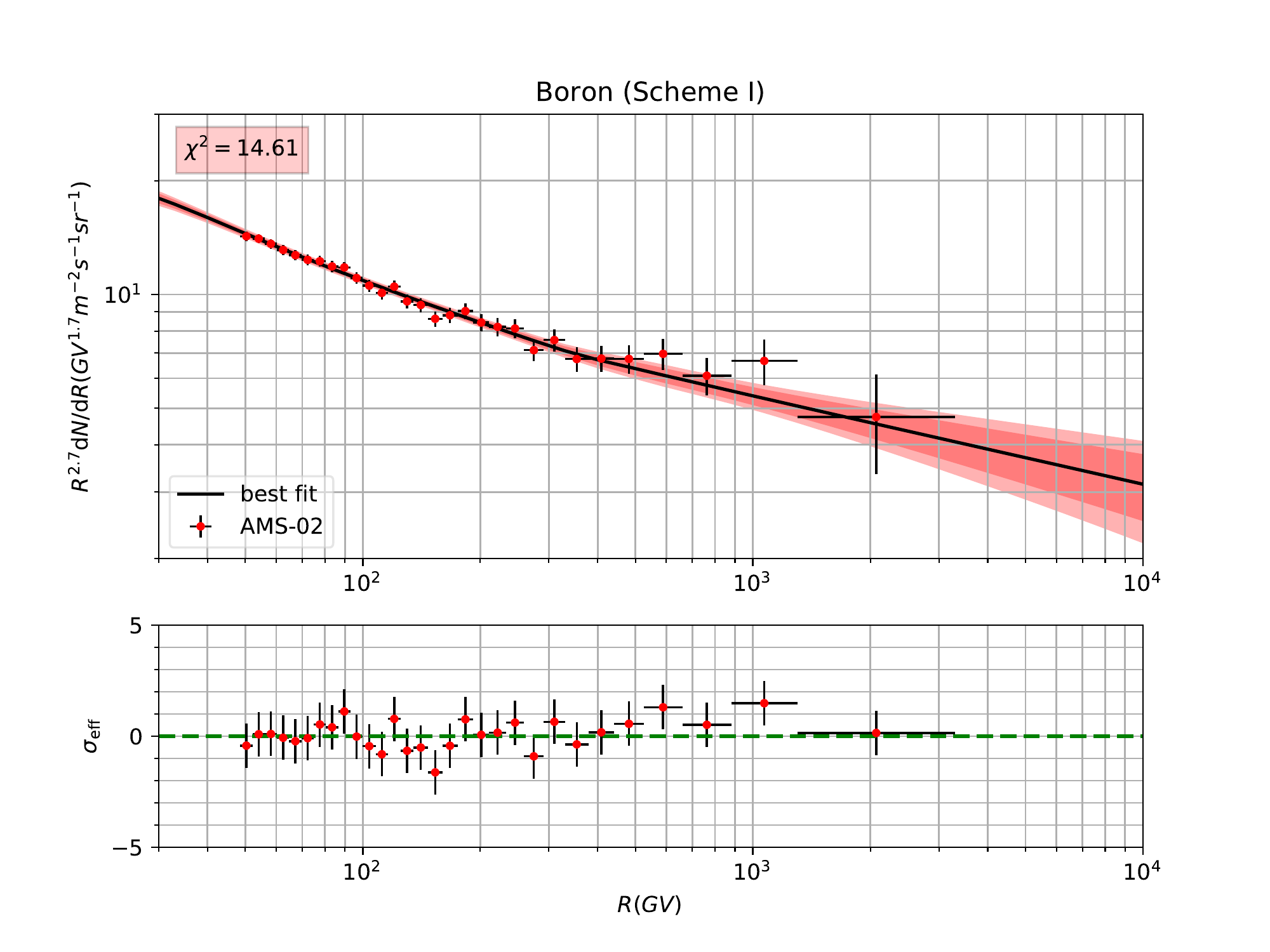}
  \includegraphics[width=0.495\textwidth]{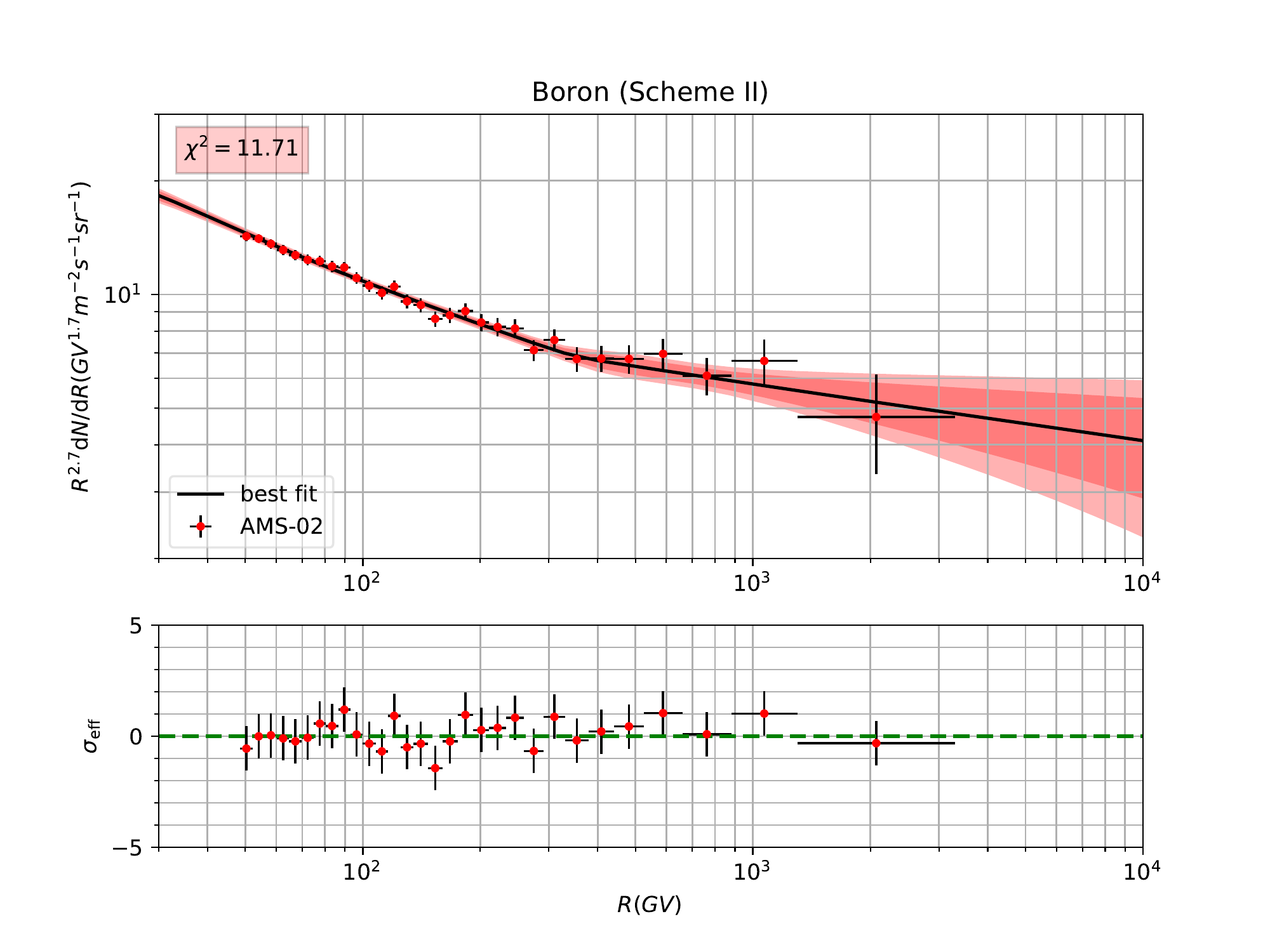}
  \caption{Same as Fig. \ref{fig:primary_results}, but for secondary nuclei species (Li, Be, and B).}
\label{fig:secondary_results}
\end{figure*}

\begin{figure*}[!htbp]
  \centering
  \includegraphics[width=0.495\textwidth]{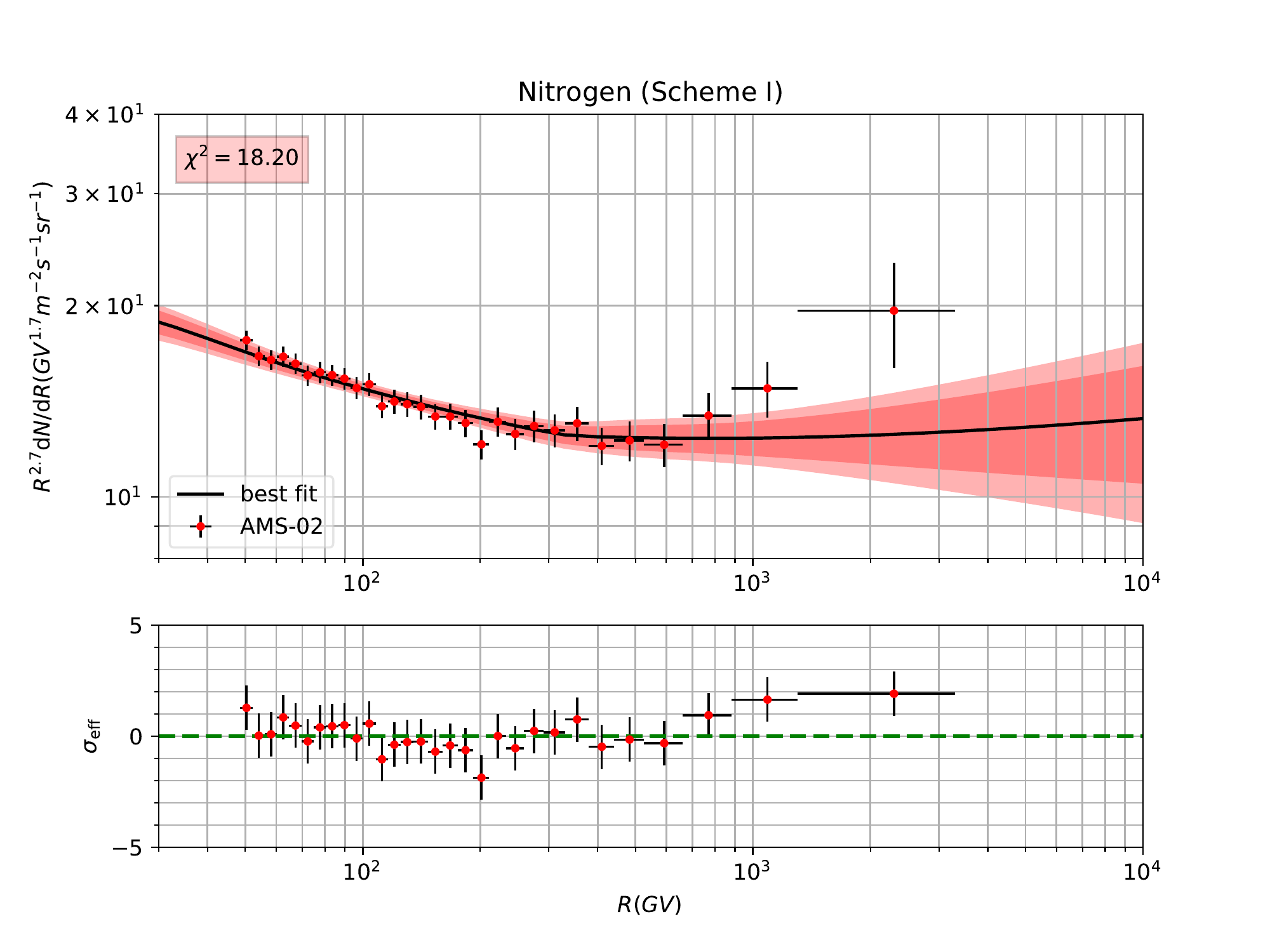}
  \includegraphics[width=0.495\textwidth]{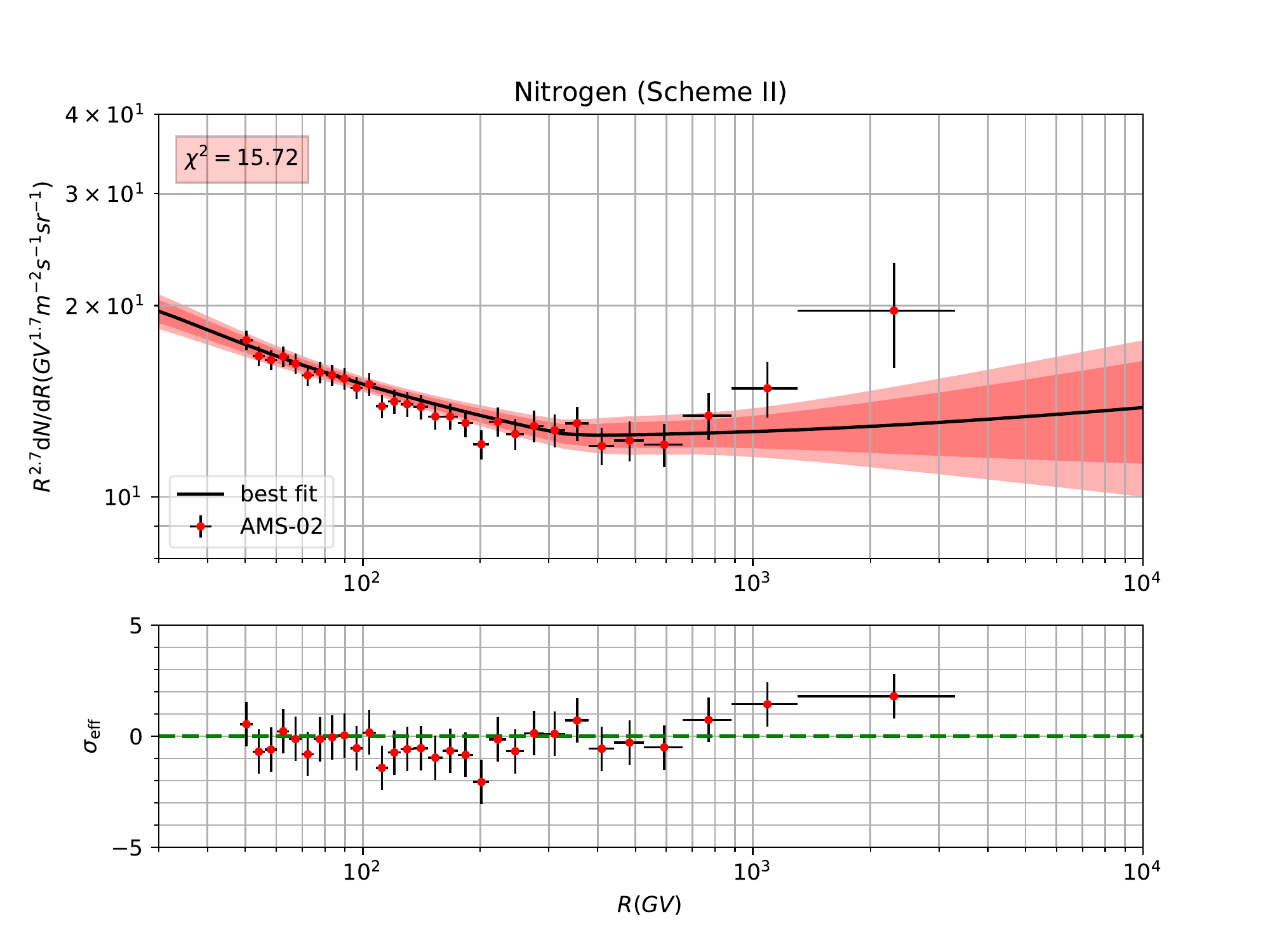}
  \caption{Same as Fig. \ref{fig:primary_results}, but for hybrid nuclei species (N).}
\label{fig:hybrid_results}
\end{figure*}

As in our previous works \citep{Niu2018,Niu2018_dampe,Niu2019_dampe,Niu201810}, the MCMC algorithm is employed to determine the posterior probability distribution of the parameters (see in Tables \ref{tab:scheme_params_I} and \ref{tab:scheme_params_I}) in Scheme I and II. The best-fit results of all the employed nuclei spectra for the two schemes are collected in Fig. \ref{fig:all_nuclei_results}. The best-fit results and the corresponding residuals (represented by $\sigeff$) of the primary nuclei species for the two schemes are given in the Fig. \ref{fig:primary_results}, that of the secondary and hybrid nuclei species are shown in Figs. \ref{fig:secondary_results} and \ref{fig:hybrid_results}. For the  best-fit results of the global fitting, we get $\chidof = 108.97/201 $ for Scheme I and $\chidof = 96.16/201 $ for Scheme II.

\begin{table*}[!htbp]
\caption{
Constraints on the parameters in set $\boldsymbol{\theta}_{1}$. The prior interval, best-fit value, statistic mean, standard deviation and the allowed range at $95\%$ CL are listed for parameters. With $\chidof = 108.97/201 $ for best-fit result.}

\begin{center}
\begin{tabular}{lllll}
  \hline\hline
ID               &Prior     & Best-fit &Posterior mean and &Posterior 95\%    \\
                 &range     &  value   &Standard deviation & range  \\
\hline
$\delta$         &[0.1, 1.0]& 0.36    & 0.36$\pm$0.01   & [0.34, 0.38]\\

$R_{\A}\ (\GV)$  &[200, 800]& 365  & 370$\pm$79& [248, 511]\\

$\nu_{\A1}$      &[1.0, 4.0]& 2.34    & 2.34$\pm$0.01   & [2.32, 2.36]\\

$\nu_{\A2}$      & [1.0, 4.0] & 2.24  & 2.23$\pm$0.03   & [2.18, 2.28]\\

\hline

$\che$           & [0.1, 5.0] & 0.655  & 0.655$\pm$0.005   & [0.646, 0.664]\\

$\cc$            & [0.1, 5.0] & 0.554  & 0.554$\pm$0.005   & [0.545, 0.562]\\

$\cnp$           & [0.1, 5.0] & 0.808  & 0.809$\pm$0.067   & [0.698, 0.923]\\

$\co$            & [0.1, 5.0] & 0.486  & 0.486$\pm$0.004   & [0.480, 0.493]\\

\hline

$\cli$           & [0.1, 5.0] & 1.94   & 1.94$\pm$0.09     & [1.79, 2.09]\\

$\cbe$           & [0.1, 5.0] & 2.28   & 2.28$\pm$0.10     & [2.12, 2.45]\\

$\cb$            & [0.1, 5.0] & 1.45   & 1.45$\pm$0.06     & [1.35, 1.56]\\

$\cns$           & [0.1, 5.0] & 1.11   & 1.11$\pm$0.10     & [0.96, 1.28]\\

\hline\hline
\end{tabular}
\end{center}
\label{tab:scheme_params_I}
\end{table*}

\begin{table*}[!htbp]
\caption{
Constraints on the parameters in set $\boldsymbol{\theta}_{2}$. The prior interval, best-fit value, statistic mean, standard deviation and the allowed range at $95\%$ CL are listed for parameters. With $\chidof = 96.16/201 $ for best-fit result.}
\begin{center}
\begin{tabular}{lllll}
  \hline\hline
ID              & Prior      & Best-fit &Posterior mean and  & Posterior 95\%    \\
                & range      &   value  &Standard deviation  & range  \\
\hline
$\Rbr\ (\GV)$   & [200, 800] & 325  & 331$\pm$70 & [233, 468]\\

$\delta_{1}$    & [0.1, 1.0] & 0.36    & 0.37$\pm$0.01    & [0.34, 0.39]\\

$\delta_{2}$    & [0.1, 1.0] & 0.26    & 0.26$\pm$0.03    & [0.21, 0.30]\\

$\nu_{\A}$      &[1.0, 4.0]  & 2.34    & 2.34$\pm$0.01    & [2.32, 2.36]\\

\hline

$\che$          & [0.1, 5.0] & 0.653    & 0.653$\pm$0.005   & [0.645, 0.661]\\

$\cc$           & [0.1, 5.0] & 0.551    & 0.551$\pm$0.005   & [0.543, 0.560]\\

$\cnp$          & [0.1, 5.0] & 0.815    & 0.805$\pm$0.067   & [0.707, 0.926]\\

$\co$           & [0.1, 5.0] & 0.479    & 0.479$\pm$0.004   & [0.473, 0.485]\\

\hline

$\cli$          & [0.1, 5.0] & 2.23     & 2.25$\pm$0.11     & [2.05, 2.40]\\

$\cbe$          & [0.1, 5.0] & 2.57     & 2.59$\pm$0.12     & [2.37, 2.76]\\

$\cb$           & [0.1, 5.0] & 1.65     & 1.67$\pm$0.08     & [1.53, 1.78]\\

$\cns$          & [0.1, 5.0] & 1.21     & 1.27$\pm$0.15     & [1.04, 1.37]\\

\hline\hline
\end{tabular}
\end{center}
\label{tab:scheme_params_II}
\end{table*}

Note that in the lower panel of sub-figures in Figs. \ref{fig:primary_results}, \ref{fig:secondary_results}, and \ref{fig:hybrid_results}, the $\sigma_{\mathrm{eff}}$ is defined as
\begin{equation}
\sigma_{\mathrm{eff}} = \frac{f_{\mathrm{obs}} - f_\mathrm{cal}}{\sqrt{\sigma_\mathrm{stat}^{2} + \sigma_\mathrm{sys}^{2}}},
\end{equation}
where $f_\mathrm{obs}$ and $f_\mathrm{cal}$ are the points which come from the observation and model calculation; $\sigma_\mathrm{stat}$ and $\sigma_\mathrm{sys}$ are the statistical and systematic standard deviations of the observed points. This quantity could clearly show us the deviations between the best-fit result and observed values at each point based on its uncertainty. Considering the correlations between different parameters, we could not get a reasonable reduced $\chi^{2}$ for each part of the data set independently. As a result, we present the $\chi^{2}$ for each part of the data set in Figs. \ref{fig:primary_results}, \ref{fig:secondary_results}, and \ref{fig:hybrid_results}.

Generally speaking, all the nuclei spectra can be well fitted in 2 schemes (the largest deviation is smaller than 3$\sigma$, see in Figs. \ref{fig:primary_results}, \ref{fig:secondary_results}, and \ref{fig:hybrid_results}.). Because the 2 schemes have a same data set and number of parameters, they have the same degree of freedom and can be compared directly by $\chi^{2}$. From the best-fit results, we get $\Delta \chi^{2} = \chi^{2}_\mathrm{I} - \chi^{2}_\mathrm{II} = 12.81$, which is a decisive evidence\footnote{In Bayesian terms, the criterion of a decisive evidence between 2 models is $\Delta \chi^{2} \geq 10$ (see, e.g., \citet{Genolini2017}). } of indicating that the current data set favors the Scheme II.
 Considering the traditional simple assumptions in Scheme I and II (assuming a simple broken power-law for injection spectra,  a uniform isotropic diffusion coefficient in the whole propagation region, etc.), we could not get a definite conclusion that the origin of the spectral hardening in nuclei spectra comes from the propagation processes, but at least it shows a tendency that current data set favors a high-rigidity break in the diffusion coefficient. More precise spectra date points in high rigidity ($\geq 2 \TV$), especially that of the secondary nuclei species, could give us more concrete conclusions.

\begin{figure*}[htbp]
  \centering
  \includegraphics[width=0.495\textwidth]{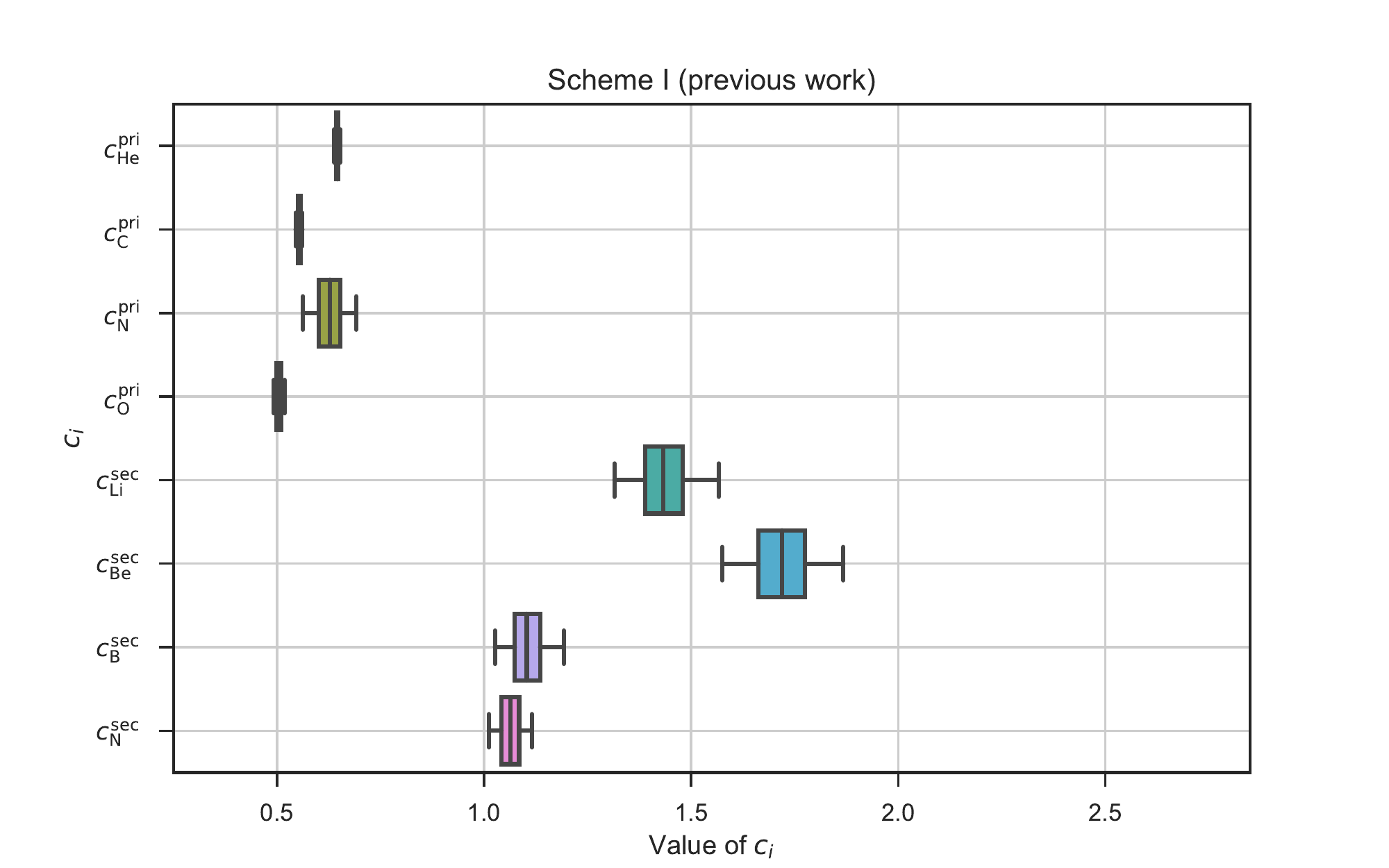}
  \includegraphics[width=0.495\textwidth]{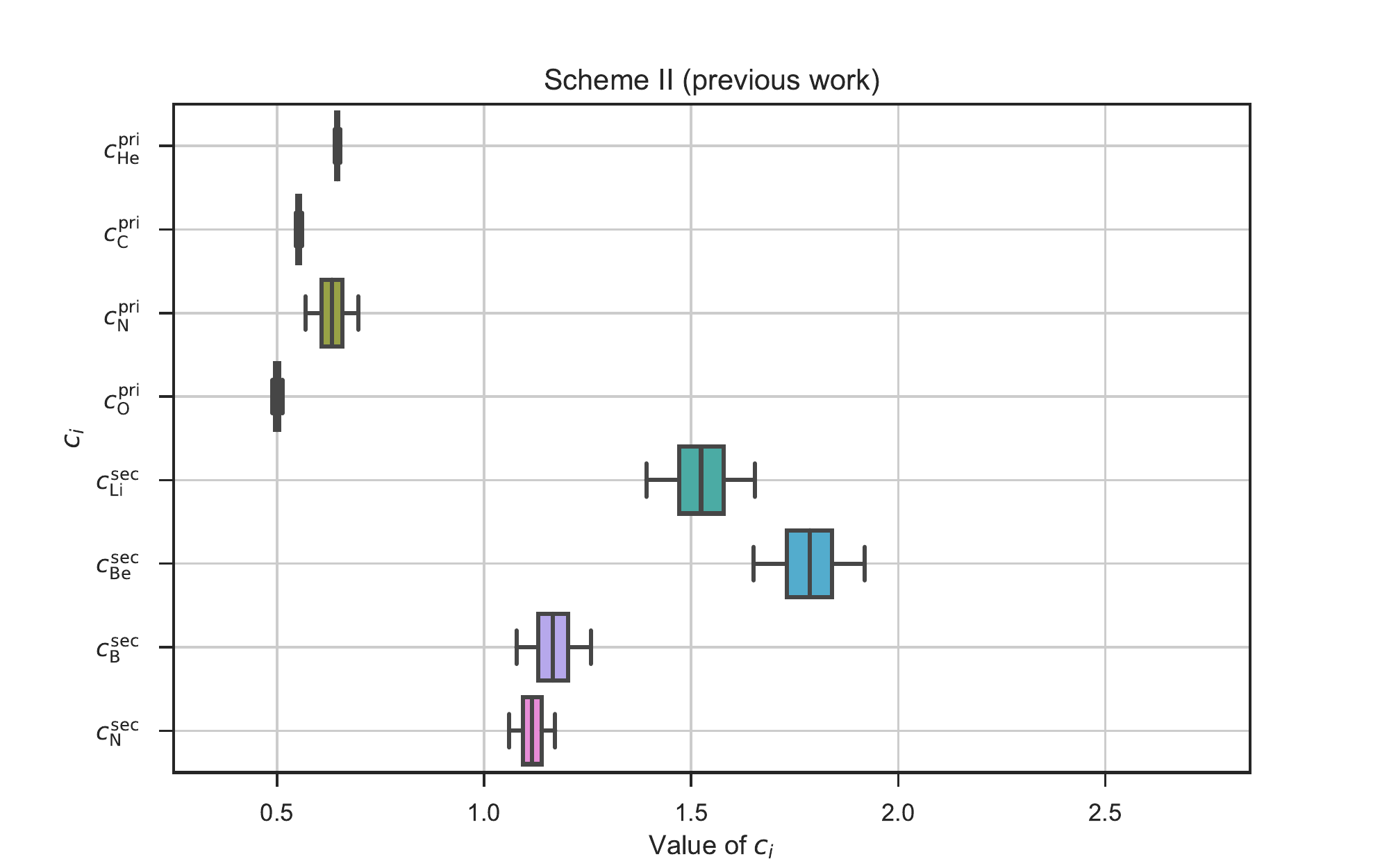}
  \includegraphics[width=0.495\textwidth]{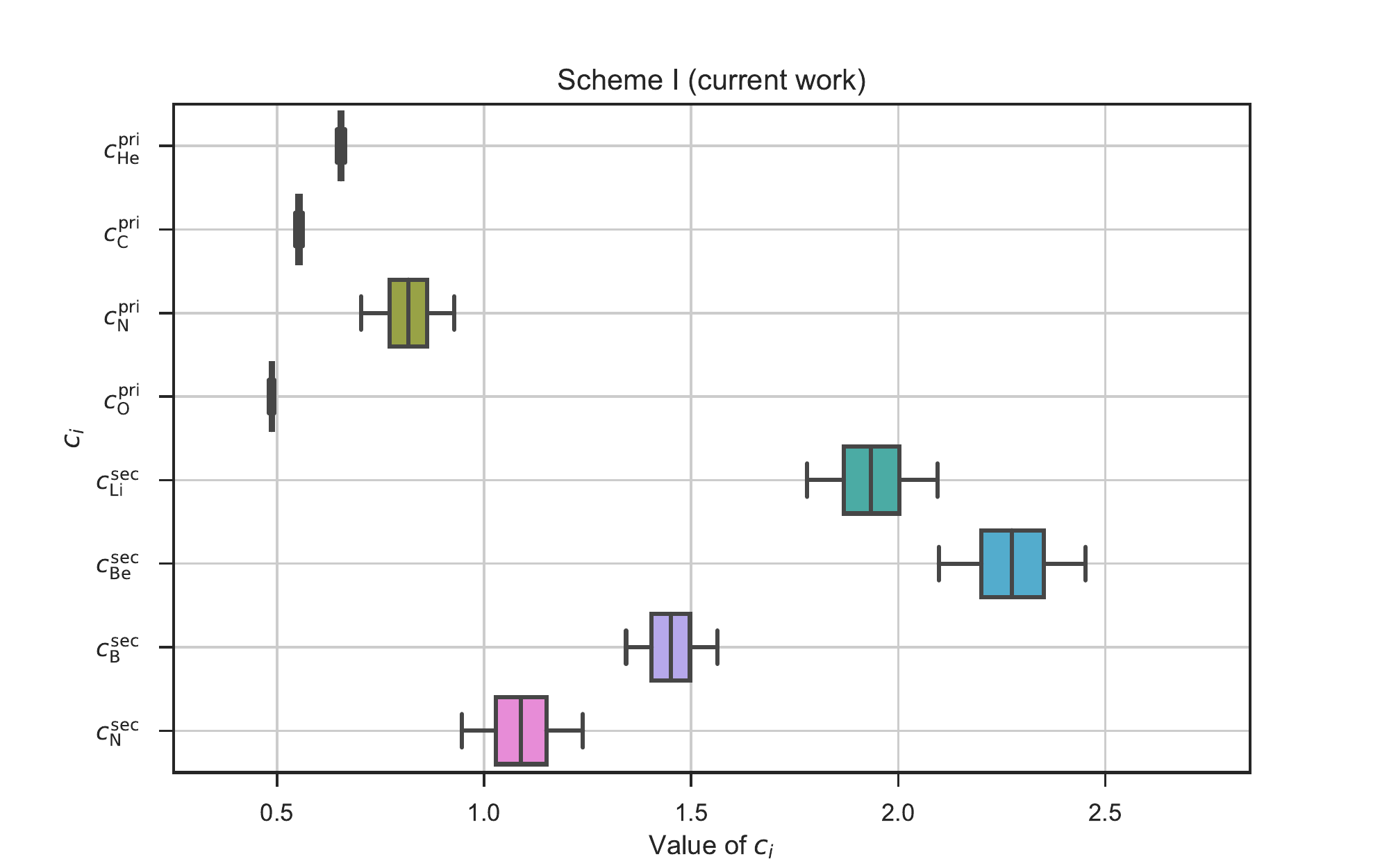}
  \includegraphics[width=0.495\textwidth]{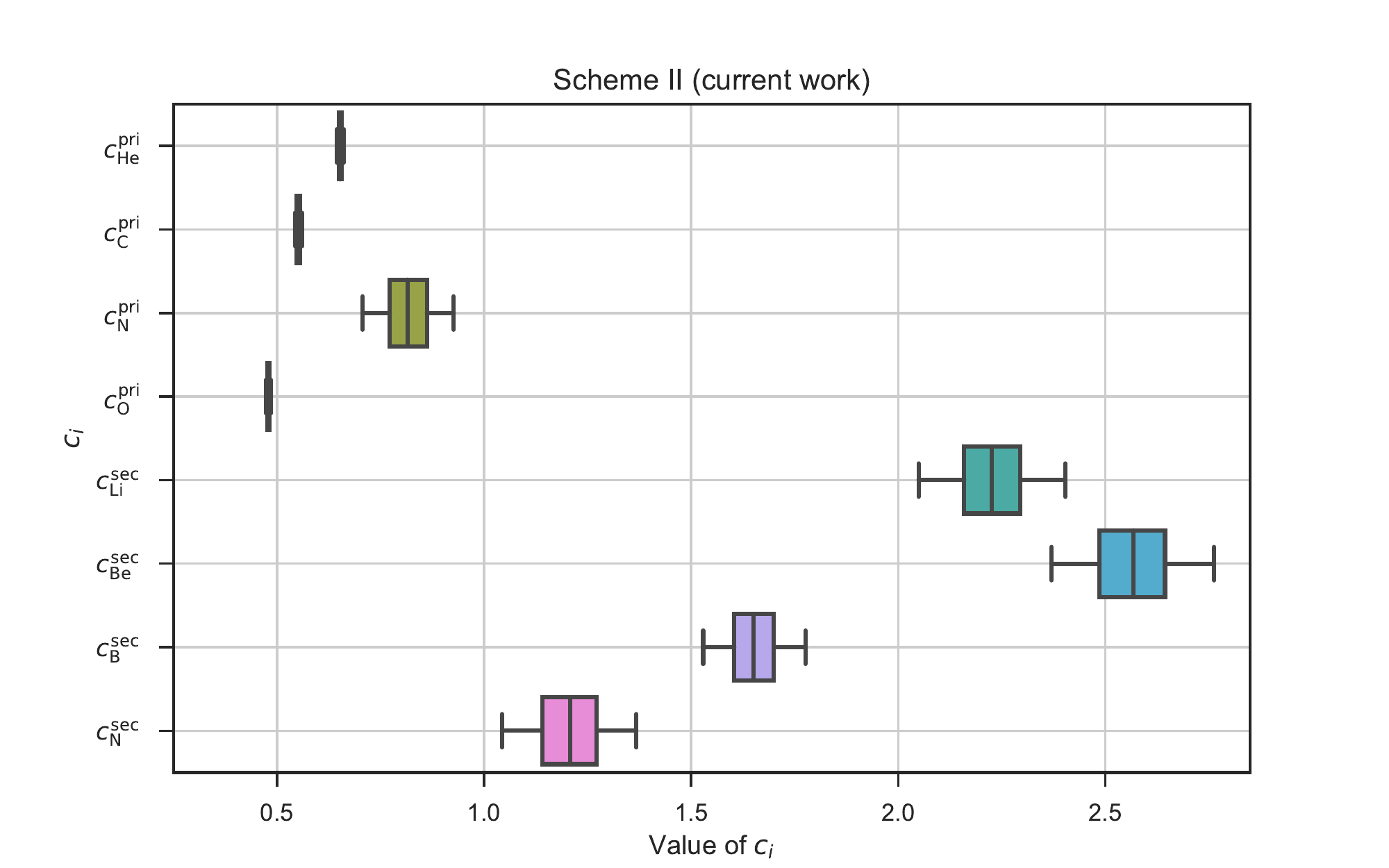}
  \caption{The boxplot for the re-scale factors of the primary and secondary components in CR nuclei species in Scheme I and II. The upper panels represent the results of our previous work which employed the AMS-02 nuclei data covering from $\sim 2 \GV$ to $\sim 2 \TV$, and the lower panels represent the results of this work which employ the AMS-02 nuclei data larger than $50 \GV$. }
\label{fig:boxplot}
\end{figure*}

The boxplot\footnote{A box plot or boxplot is a method for graphically depicting groups of numerical data through their quartiles. In our configurations, the band inside the box shows the median value of the dataset, the box shows the quartiles, and the whiskers extend to show the rest of the distribution which are edged by the 5th percentile and the 95th percentile.} for the $c_{i}$s in this work are shown in the lower panels of Fig. \ref{fig:boxplot}. For comparison, the corresponding results of our previous work \citep{Niu201810}, in which the entire AMS-02 nuclei data (including the data points $<$ 50 GV) is used in the global fitting, are shown in the upper panels in Fig. \ref{fig:boxplot}.
 We want to emphasize that in both of these works,  all the nuclei spectra are considered in a self-consistent way and all of them  are related to each other intrinsically. The fitting results clearly show that we could not reproduce the spectra of secondary species self-consistently without the employment of $\csec$s. Consequently, all the fitting results of $\cpri$s and $\csec$s should be taken seriously.
In Fig. \ref{fig:boxplot}, it is clearly shown that, same as that in previous work, no matter in Scheme I or II, the values of $\cpri$s are systematically smaller than 1.0, while the values of $\csec$s are systematically larger than 1.0.
Moreover, the values of $\cpri$s in this work are almost the same as that in previous work if we have considered the fitting uncertaintes, while the values of $\csec$s in this work are systematically larger than that in previous work.

As the nuclear charge number increases, both $\cpri$s and $\csec$s have smaller values except $\cnp$ and $\cbe$, respectively. Because the CR spectrum of nitrogen is composed by both primary and secondary components and has relative large fitting uncertainties, we will not focus on the value of $\cnp$ in this work. From the point view of $\csec$s, beryllium is the most special CR secondary species.

\section{Discussion and Conclusion}

In this work, we considered 2 schemes to reproduce the newly released AMS-02 nuclei spectra (He, C, N, O, Li, Be, and B) when $R > 50 \GV$. The fitting results show that current data set favors a high-rigidity break at $\sim 325 \GV$ in diffusion coefficient rather than a break at $\sim 365 \GV$ in primary source injections, which is consistent with the results obtained in \citet{Genolini2017}.
 Moreover, the fitted values of $\cpri$s (which are the factors to rescale the default isotopic abundances of helium-4, carbon-12, nitrogen-14, and oxygen-16 in {\sc galprop}) are systematically smaller than 1.0 and consistent with the results in our previous work \citep{Niu201810} within fitting uncertainties. While the fitted values of $\csec$s (which are the factors to rescale the CR flux of secondary species/components after propagation) are systematically larger than 1.0 and larger than the values obtained in our previous work \citep{Niu201810}, which includes the entire spectra data points in the global fitting.

 In some of the previous works (see, e.g., \citet{Lin2015,Lin2017,Yuan2017,Niu2018,Niu2018_dampe,Niu2019_dampe}), the $\pbar$ rescale factor $\cpbar$ always have a value of $\sim 1.3$, which has been explained to approximate the ratio of antineutron-to-antiproton production cross section \citep{Mauro2014}. Similarly, all the other $\csec$s could be interpreted as the same origins. However, generally speaking, the production cross sections of these secondary species are energy dependent. In Figs. \ref{fig:all_nuclei_results}, \ref{fig:primary_results}, \ref{fig:secondary_results}, and \ref{fig:hybrid_results}, one can find that the fitting results are quit well in most of the cases. If this is the right explanation, it is quit unnatural that all these secondary species have energy independent corrections on their production cross sections. On the other hand, it is also quit unnatural that all the production cross sections of these secondary species have been underestimated simultaneously. As a result, this explanation could be excluded to some extend. At least, it could not be the dominate factor.

 According to observing the extended emission around Geminga and PSR B0656+14 pulsar wind nebulae (PWNe), \citet{HAWC2017} have found that the estimated CR diffusion coefficient ($D_{0}$) are more than two orders of magnitude smaller than the typical value derived from the secondary/primary nuclei species in galactic CRs. It infers that there exists slower diffusion zone (SDZ) around PWN, which could be extended to CR sources \citep{Johannesson2019}. Some previous works \citep{Fang2018,Fang2019apj} show that positron excess in CRs could be explained by this two-zone model. If we locate in such a SDZ \citep{Fang2019mn}, a smaller $D_{0}$ can lead to produce more secondary nuclei species' flux than the uniform diffusion in the whole galaxy. This could be the solution to the underestimation of secondary flux in current model.

 Meanwhile, the $\csec$s in current work (averaged from 50 GV to 2 TV ) are systematically larger than that in our previous work (averaged from 2 GV to 2 TV), which implies that it needs more secondary CR particles in high energy region to meet the observed data.\footnote{Of course, it might come from the imperfect model on solar modulation in our previous work in low energy region ($R < 50 \GV$).}  If it is the case, one needs extra injection of secondary nuclei species in high energy region. This scenario is recently studied by some interesting works (see, e.g., \citet{Yang2019} and \citet{Boschini2019}).

 In summary, we could ascribe the underestimation of the CR secondary flux to the SDZ which we locate in. At the same time, another hint implies it might need extra injection of secondary CR particles in high energy region. All the related details need more attention in future research.

\section*{ACKNOWLEDGMENTS}
Jia-Shu Niu would like to appreciate Yi-Hang Nie and Jiu-Qin Liang for their trust and support.
This research was supported by the Special Funds for Theoretical Physics in National Natural Science Foundation of China (NSFC) (No. 11947125) and the Applied Basic Research Programs of Natural Science Foundation of Shanxi Province (No. 201901D111043).

%

\end{document}